\documentstyle[11pt,psfig]{article}

\input{epsf}

\begin{document}

\sloppy
\setcounter{page}{0}
\thispagestyle{empty}

\begin{flushright}
nucl-th/9706002
\end{flushright}
\vspace{3.5cm}

\begin{center}
{\Large
{\bf PARTON CASCADE DESCRIPTION 
\\${}$\\
OF RELATIVISTIC HEAVY ION COLLISIONS
\\${}$\\
AT {\sl CERN SPS} ENERGIES  \Huge{?}}
}
\end{center}
\bigskip

\begin{center}
{\Large {\bf Klaus Geiger$^1$ and Dinesh Kumar Srivastava$^2$}}
\vskip 0.2in
$^1${\large{\em Physics Department, Brookhaven National Laboratory,
Upton, N. Y. 11973, U. S. A.}}
\\
$^2${\large{\em Variable Energy Cyclotron Centre, 1/AF Bidhan Nagar, Calcutta
700 064, India}}
\end{center}
\vskip 0.5in
\begin{center}
{\bf Abstract}
\end{center}
\vskip 0.2in
We examine $Pb+Pb$ collisions at CERN SPS energy 158 $A$ GeV,
by employing the earlier developed and recently refined
parton-cascade/cluster-hadronization model
and its Monte Carlo implementation.
This space-time model
involves the dynamical interplay of perturbative QCD
parton production and evolution, with non-perturbative  parton-cluster
formation and hadron production through cluster decays.
Using computer simulations, we
are able to follow
the entwined time-evolution of parton and hadron degrees of freedom 
in both position and momentum space, from the instant of nuclear overlap to the
final yield of particles.
We present and discuss results
for the multiplicity distributions, which agree well
with the measured data from the CERN SPS, including those for K mesons. The
transverse momentum distributions of the produced hadrons are also found to
be in good agreement with the preliminary data measured by the NA49 and the
WA98 collaboration for the collision of lead nuclei at the CERN SPS.
The analysis of the time evolution of transverse energy deposited in
the collision zone and the energy density suggests an existence of
partonic matter for a time of more than 5 $fm$. 
\vskip 0.5in
\leftline{{\normalsize PACS numbers: 12.38.Bx, 12.38.Mh, 25.75.+r, 24.85.+p}}

\newpage

\section{INTRODUCTION}
\label{sec:section1}
\bigskip

The study of relativistic heavy ion collisions over the last several years
has been driven by the motivation to 
discover novel phenomena associated with the collective behavior of
highly compressed QCD matter. At the forefront of this effort
lies the search for the `notorious' 
quark-gluon plasma (QGP), a thermal state of colored partons deconfined 
over a macroscopic volume.
A large number of experiments
have already been conducted at the CERN SPS with oxygen, sulphur, and lead
beams. Some of the proposed signals like a suppression of J/$\Psi$
production, an enhanced production of strangeness, and an excess production
of dileptons have already been observed. 

A large number of papers~\cite{review}
in the literature explicitly start with an {\it assumption} of the
actual existence of a QGP at some stage during the nuclear collision.
Leaving aside the most interesting question whether and how such a QGP
may be realistically formed through the evolution of the initial nuclear
state,
further assumptions about the specific quark-gluon composition,
volume, density and temperature, etc.,
are usually employed due to the lack of a better knowledge.
Are these assumptions too ``bold'' or ``tantalizing''? 
What is the justification for such assumptions
except for the plea that hadronic densities may get too large for such
collisions, and thus a treatment in terms of hadrons may get unreasonable?
Is this plea too specious? Also, have such collisions succeeded in creating
a reasonably large volume over which the energy density is large enough to
admit a QCD phase transition?

Quite a few models have been used to 
describe the gross features of these collisions in 
terms of the so-called {\it string picture} for hadronic interactions, 
i.e., based on  modeling  nuclear
collisions in terms of nucleon-nucleon collisions on the
basis of a constituent valence-quark picture plus
string-excitation and -fragmentation.
Popular examples for these models are
FRITIOF \cite{fritiof}, 
VENUS \cite{venus}, 
RQMD \cite{rqmd},
DPM \cite{dpm}.
The continuous refinements and the fine tunings  of the necessarily 
involved parameters 
have resulted in development of several generations of these `event generators'. 
In particular VENUS and RQMD are  impressively fine-tuned to describe the recent
CERN SPS data. Yet, the price paid is the invention
of supplemental mechanisms (such as `string droplets' in VENUS or `color ropes' in RQMD)
to mimic certain  underlying dynamics due to nuclear effects, which
cannot be accounted for in  a non-trivial manner in these models.
What does the success of VENUS and RQMD
imply, in view of the fact that the `strings' and the `ropes' only 
serve to mimic the
actual interactions? These studies have also lead to a  ``faith'' that the
interactions at the SPS energies is mostly ``soft''. Should we remain 
complacent in this belief?

At the same time,
a large body of hadronic data measured at the SPS energies has been 
shown~\cite{johanna} to
be fairly consistent with a {\it thermal model}, suggesting a thermal and
chemical equilibrium among hadrons at the time of freeze-out.
A purely hadronic picture of the interaction is not very likely to drive
the system to the verge of chemical equilibrium~\cite{satz}, as the hadronic 
reactions are far too slow.

The provocative question we dare to pose then is:
How can we really improve the knowledge of the underlying
properties of QCD in dense matter, when  drastically
different phenomenological approaches (with their
very own parametrizations of unknown physics details)
 describe the experimental data more or less equally well?

We are convinced that this issue and many of the above-mentioned
questions can be settled satisfactorily in the near future,
 if we attempt to understand the collision among
heavy ions at these energies at a more microscopic level by
borrowing and extending rigorous QCD techniques that have been
developed over the years in particle physics.
Clearly, the entwined dynamics of partonic and
hadronic degrees of freedom that we have to face,
requires that we consider a mixed 
parton-hadron system with its space-time dependent relative proportions being
determined by the multi-particle dynamics itself.

We  therefore advocate that  
for ultra-relativistic nucleus-nucleus collisions
a description based on
the pQCD interactions and cascade evolution of involved partons 
can and should be used, owing to the claim that
short-range parton interactions play 
(at least during the early and most dissipative stage of the first few $fm$) 
an important role at  sufficiently high beam energies
(say, $\sqrt{s}  \, \lower3pt\hbox{$\buildrel >\over\sim$}\,20$ A GeV).
Here copiously produced quark-gluon mini-jets cannot be considered as 
isolated rare events, but must be embedded in the complex multiple 
cascade-type processes.
The present study is a  step in this direction.

We advertize
an elaborate QCD-based space-time model that allows to simulate 
nucleus-nucleus collisions (among other particle collisions), 
which is now available 
\footnote{
The program VNI-3.1 (pronounced {\it Vinnie}) is available
from http://rhic.phys.columbia.edu/vni.
}
as a computer simulation program called VNI \cite{ms44}.
It is a pQCD parton cascade description \cite{msrep}, 
supplemented by a phenomenological
hadronization model \cite{EG},
with dynamically changing proportions of partons and hadrons, in
which the evolution of a nuclear collision is traced 
from the first instant of overlap, via QCD parton-cascade development
at the early stage,
parton conversion into pre-hadronic color-singlet clusters and hadron
production through the decays of the clusters, 
as well as the fragmentation of the beam remnants at late times. 
(We will in the following refer to the parton-cascade/cluster-hadronization
 model as PCM, and to its Monte Carlo implementation as VNI.)
The PCM description has been
used \cite{msrep} to provide very useful insight into the dynamics of
 the evolution of the matter at
energies likely to be reached at BNL RHIC and CERN LHC. The experimental
data for these will, however, come only in the next millennium.

Returning to our above-posted hypothesis of the applicability
of the pQCD parton-picture for high-energy
nuclear collisions, we must address the question:
{\it
Can we use the PCM picture already at the SPS energies ($\sqrt{s}\approx$ 17--20
GeV/A)?
}
Recall that the parameters of the models were fixed by the
experimental data for 
$pp$ ($p\overline{p}$) cross-sections over $\sqrt{s}$ = 10--1800 GeV, and
$e^+e^-$ annihilation \cite{msrep}. The nucleon-nucleon energy reached at SPS is
well within this range.
Indeed, as we shall show, this approach does remarkably well in comparison
to the gross particle production properties observed at CERN SPS.
This comes as a surprise to us, since
no attempt was made to fine-tune the model to the data.
In view of this success, a very good opportunity
opens to understand
these collisions in their entirety, where the early stages are fairly well
understood in terms of the pQCD. This will allow us to draw
definite conclusions about the initial energy density and the constitution of
the matter produced. This will also give us confidence that projections
for the RHIC and LHC energies using the PCM may be reasonable. 
\medskip

In Sect. 2 we briefly describe the PCM, and its
Monte Carlo implementation VNI. In Sect. 3, we first discuss the general outcome
of a model study for central collisions of lead nuclei as SPS energies.
In particular we explore the contribution of various processes to the
transverse energies deposited in the collision
and examine the time-evolution and magnitude of the energy
density in the central collision region. Next we present our results
for the rapidity distributions for the transverse energy, multiplicities,
and the transverse momentum spectra of hadrons and compare them with the
available experimental data. We find a good agreement and discuss the
consequences.  Finally, Sect. 4 gives a summary.
\bigskip
\bigskip

\section{ASPECTS OF THE MODEL}
\label{sec:section2}
\bigskip

Let us recall the main aspects of the PCM, which
are specific to our analysis of heavy-ion collisions afterwards.
For the sake of brevity, we will not review the
model in detail, but kindly refer
to the extensive documentations of Refs.
\cite{msrep,ms44}.
\medskip

\subsection{Theoretical framework}
\smallskip

The central element in the
space-time cascade description is the use of
relativistic transport theory \cite{msrep} in conjunction
with renormalization-group  improved QCD \cite{ms39},
which provides the theoretical basis
to follow the QCD evolution 
in 7-dimensional phase-space $d^3 r d^3 k dE$
of a mixed 
multi-particle system of partons and hadrons
with dynamically changing proportions.
We remark, that
a  theoretical basis for
such a space-time cascade description of a multiparticle system
in high-energy collisions can be derived systematically from
{\it quantum-kinetic theory} on the basis of
QCD's first principles in a stepwise approximation scheme 
(see  e.g., Refs. \cite{ms39} and references therein).
This framework allows  to
cast the time evolution of the mixed system of
individual partons, composite parton-clusters, and physical hadrons
in terms of a closed set of
integro-differential equations for
the phase-space densities of the different particle excitations.
The definition of these phase-space densities,
denoted by
$F_\alpha$, where $\alpha\equiv p, c, h$
labels the species of partons, pre-hadronic clusters, or hadrons,
respectively, is:
\begin{equation}
F_\alpha(r,k)\;\,\equiv\; \, F_\alpha (t, {\bf r}; E, {\bf k})
\;\,=\;\,
\frac{dN_\alpha}{d^3r d^3k dE}
\;\;\;\;\;\;\;\;
(\alpha\;\equiv\; p, \;c, \;h)
\;,
\label{F}
\end{equation}
where 
$r\equiv r^\mu=(t,{\bf r})$, $k\equiv k^\mu=(E,{\bf k})$,
and $k^2 =k_\mu k^\mu = E^2 -{\bf k}^{\,2}$ can be off-shell 
(space-like $k^2 < m^2$, time-like $k^2 > m^2$) or on-shell
 ($k^2 = m^2$).
The densities (\ref{F}) measure the number of particles
of type $\alpha$ at time $t$ with position in ${\bf r} + d{\bf r}$,
momentum in ${\bf k} + d{\bf k}$,
and energy in $E + dE$ (or equivalently invariant mass in 
$k^2 + dk^2$).
The $F_\alpha$ are the quantum analogues of the
classical phase-space distributions, including both off-shell 
and on-shell
particles, and hence
contain the essential microscopic
information required for a statistical description
of the time evolution of a many-particle system in
complete 7-dimensional phase-space $d^3rd^3kdE$, 
thereby providing the basis for calculating
macroscopic observables.

The phase-space densities (\ref{F}) are determined by the
self-consistent solutions of
a set of {\it transport equations} 
(governing the space-time change with $r^\mu$)
coupled with
renormalization-group-type {\it evolution equations}
(controlling the change with momentum scale $k^\mu$)
\cite{ms39}.
These equations can be generically expressed as
convolutions of the densities $F_\alpha$ of particle species
 $\alpha$,
interacting with  specific cross-sections $\hat{\sigma}^{(l)}$ 
for the processes $l$.
The resulting coupled equations for the
space-time development of
the densities of partons $F_{p}$, clusters $F_c$ and
hadrons $F_h$ is a self-consistent set in which the change
of the three distinct densities is governed by the balance of
the various possible interaction processes among the particles.
The generic form is
\begin{eqnarray}
k_\mu \frac{\partial}{\partial r^\mu}\; F_\alpha(r,k)
&=&
\sum_{\beta,\ldots}\,\sum_l\; 
\left\{
\frac{}{}
\hat{I}_{gain}^{(l)}[\hat{\sigma}^{(l)},F_\alpha,F_\beta,\ldots ]
\;-\;
\hat{I}_{loss}^{(l)}[\hat{\sigma}^{(l)},F_\alpha,F_\beta,\ldots ]
\right\}
\label{e1}
\\
k^2  \frac{\partial}{\partial k^2}\; F_\alpha(r,k)
&=&
\sum_{\beta,\ldots}\,\sum_l\; 
\left\{
\frac{}{}
\hat{J}_{gain}^{(l)}[\hat{\sigma}^{(l)},F_\alpha,F_\beta,\ldots ]
\;-\;
\hat{J}_{loss}^{(l)}[\hat{\sigma}^{(l)},F_\alpha,F_\beta,\ldots ]
\right\}
\label{e2}
\;,
\end{eqnarray}
where $\alpha, \beta, \ldots\;\equiv\; p, \;c, \;h$.
Equations (\ref{e1}) and (\ref{e2}) allow a {\it probabilistic
interpretation} of the multi-particle evolution in 
space-time and momentum space
in terms of sequentially-ordered interaction processes $l$,
in which the rate of change of the particle distributions 
$F_\alpha$
($\alpha=p,c,h$)
in a phase-space element $d^3rd^4k$
is governed by the balance of gain (+) and loss ($-$) terms.
The left-hand side
describes free propagation of a
quantum of species $\alpha$, whereas
on the right-hand side the interaction kernels $\hat{I}$, 
$\hat{J}$
are integral operators that incorporate the effects of
the particles' self-  and mutual interactions.
Explicit expressions are given in Refs.  \cite{msrep,ms37}.
Here we merely note that the kernels $\hat{I}$ and $\hat{J}$
 embody
convolutions of
the density of particles $F_{\alpha},F_\beta,\ldots$ 
entering or leaving  a particular vertex,
and a phase-space integration
weighted with the associated  probability distribution,
i.e. with the relevant cross-section $\hat{\sigma}^{(l)}$,
of the squared amplitude. 
\medskip

\subsection{Practical application}
\smallskip

In practice, the above formalism allows one to
trace the microscopic history of the dynamically-evolving 
particle system
in space-time {\it and} momentum space, so that
the correlations of particles in space,  time, color and flavor
 can be taken
into account systematically.
The interplay
between perturbative and non-perturbative regimes is controlled
 locally
by the space-time evolution of the mixed parton-cluster-hadron
 system itself
(i.e., the time-dependent local particle densities).

As mentioned, these concepts are implemented in the Monte Carlo 
program VNI \cite{ms44}
which simulates high-energy particle collisions
on the basis of the  PCM concepts.
For nucleus-nucleus collisions,
there are 
three main building-blocks, describing the  collision dynamics
from the initial  beam/target collision system
upon collisional  contact, through the QCD-evolution of
parton distributions, hadron formation,
up to the emergence of final hadronic states:
\begin{description}
\item[(i)]
The {\it initial state} associated with the incoming
nuclei involves their decomposition into nucleons, and, of the 
nucleons into
partons on the basis of the experimentally measured  
nucleon structure functions and elastic form-factors.
This procedure translates the initial nucleus-nucleus system into
two colliding clouds of {\it virtual} partons according
to the well-established parton decomposition of
the nuclear wave functions at high energy \cite{GLR}.
\item[(ii)]
The {\it parton cascade development}
starts from the initial interpenetrating parton clouds,
and involving  the space-time
development with mutual- and self-interactions of the 
system of quarks and gluons.
Included are multiple elastic and inelastic interaction processes,
 described 
as sequences of elementary $2 \rightarrow 2$ scatterings,
 $1\rightarrow 2$
emissions and $2 \rightarrow 1$ fusions.
Moreover,  correlations are accounted for between primary virtual
partons, emerging as unscathed remainders from the initial 
state, and
secondary real partons, materialized or  produced 
through the partonic interactions.
\item[(iii)]
The {\it hadronization dynamics} of the evolving system
in terms of parton-coalescence to color-neutral clusters
is described as a local, statistical process that depends on the 
spatial separation
and color of nearest-neighbor partons.
Each pre-hadronic parton-cluster fragments through isotropic
 two-body decay
into  primary hadrons, according to the density of
states, followed by the decay of the latter into final 
stable hadrons.
\end{description}
\medskip

\subsection{Simulation procedure}
\smallskip

The simulation of the time development of the mixed system
of partons, clusters, and hadrons
in position and momentum space on the basis of 
eqs. (\ref{e1})-(\ref{e2})
emerges then from following each individual particle through its 
history with
the various  probabilities and time scales of interactions 
sampled stochastically from the relevant probability distributions
in the kernels $\hat{I}$ and $\hat{J}$.
The microscopic history of the system can thus be traced by evolving
the phase-space distributions of particles are
evolved in small time steps ($\Delta t \simeq 10^{-3}\;fm$)
 and 7-dimensional phase-space $d^3rd^3kdE$ throughout 
the stages of parton cascade, parton-cluster formation, cluster-hadron
 decays,
until stable final-state hadrons and other particles
are left as freely-streaming particles.
The essential ingredients in this Monte-Carlo procedure are summarized 
as follows
\cite{ms44}:
\begin{description}
\item[(i)]
The {\it initial state} is constructed in three steps. First, the nuclei
are decomposed into the nucleons with an appropriate
Fermi-distribution. Second, the nucleons are in turn decomposed into
their parton substructure according to proton/neutron structure functions
with a spatial distribution given by the Fourier transform 
of the nucleon elastic form-factor.
Third, the so-initialized phase-space densities of
(off-shell) partons are then boosted with
the proper Lorentz factor to the center-of-mass frame of the
colliding nuclei.
\item[(ii)]
The {\it parton cascade} development
proceeds then by
propagating the partons along classical trajectories until they interact,
i.e., collide (scattering or fusion process),
decay (emission process) 
or coalesce to pre-hadronic composite
states (cluster formation). Both space-like and time-like radiative
corrections are included within the Leading-Log approximation.
The relevant interaction probabilities, entering in the kernels $\hat{I}$, 
$\hat{J}$
in (\ref{e1}) and (\ref{e2}), are obtained from the 
well-known perturbative QCD cross-sections,
and
the coalescence probabilities of the Ellis-Geiger model \cite{EG},
respectively.
Both the production of partons and the emergence of pre-hadronic clusters
through their coalescence are subject to an individually specific 
formation time
$\Delta t_{p,c} = \gamma / M_{p,c}$ where $1/M_{p,c}=1/\sqrt{k^2}$ is the 
proper
decay time of off-shell partons or clusters with invariant mass $M_p$,
respectively  $M_c$, and $\gamma=E/M_{p,c}$ is the Lorentz factor.
\item[(iii)]
The {\it cluster-decay} and {\it hadron formation};
the decay of the pre-hadronic clusters
and the decays of excited hadrons and resonances which are 
sampled from the particle data tables \cite{pada}.
Again, each newly produced hadron becomes a `real' particle only after
a characteristic formation time 
$\Delta t_{h} = \gamma / M_{h}$ depending on their invariant mass $M_h$ and
their energy through $\gamma= E/M_h$. Before that time has passed, a hadron
must be considered as a still virtual object that cannot interact
incoherently until it has formed according to the uncertainty principle.
\item[(iv)]
The {\it beam remnants}, being the unscathed remainders of the initial nuclei, 
emerge  from reassembling all those  remnant primary partons that
have been spectators without interactions throughout the evolution.
The recollection of those yields two corresponding beam clusters
with definite charge, baryon number, energy-momentum and position, as given by
the sum of their constituents.
These beam clusters  decay into final-state hadrons which recede
along the beam direction at large rapidities of the beam/target fragmentation 
regions.
Again, individual formation times of the produced hadrons are accounted for. 
\end{description}
We note that the spatial density and the momentum distribution
of the particles are intimately connected: The momentum 
distribution continuously changes through the interactions and
determines how the quanta propagate in coordinate space.
In turn, the probability for subsequent interactions depends on the 
resulting local particle density. Consequently, the development
of the phase-space densities is a complex
interplay, which - at a given point of time - contains implicitly the
complete preceding history of the system.
\bigskip
\bigskip

\section{RESULTS FOR $Pb+Pb$ COLLISIONS AT 158 A GeV}
\bigskip

We apply now the PCM to collisions involving lead nuclei 
 ($E_{\mbox{beam}} =158 \cdot A$ GeV)
studied in detail at the CERN SPS by a number of experiments.
It is important to stress again, that
no attempt was made to fine-tune the model to the data, that is,
we used VNI with all its default settings, which
are based on $e^+e^-$ and $pp$ ($p\bar{p}$) physics \cite{msrep}.
\medskip

\subsection{`Hard' versus `soft' physics}
\smallskip

The creation of a deconfined strongly interacting matter is crucially
dependent on the energy deposited in the interaction zone by the
colliding nuclei. We have plotted the transverse energy distribution
for a central collision in Fig.~1. The individual
contributions to the transverse
energy from parton cascades and the beam fragmentation are self-evident.
We see that at the SPS energies, the `beam remnants',
i.e., the primary partons which remained spectators
throughout the evolution  account for about 50\% of the
transverse energy, at central rapidities. This partonic fraction
is seen to damp out quickly as we move away from the central rapidity
and approach the beam/target rapidities. We have additionally estimated
the contribution of multiple scatterings among partons by switching
off all but the collision between the primary partons. This reduced the
$dE_T/dy$ at $y=0$ by about 10\%, implying that up to 20\% of the energy
deposited in partonic collisions originates from multiple scatterings.
This will certainly alter at higher energies. 
\smallskip

We emphasize that the pQCD
parton cascading is treated strictly within perturbative QCD,
whereas the non-perturbative hadronization dynamics
is a phenomenological prescription that has been shown
to work equally  well for $e^+e^-$, $pp$ and AA collisions  \cite{msrep}.
The rationale is 
keeping these two elements rigorously separate and not entangling them
in a foggy manner.
However, there is an additional
important aspect which is inherent to any pQCD description of parton evolution,
namely,
the well-known defect that - for massless quarks and gluons - the
interaction integrals $\hat{I}^{(l)}$ and $\hat{J}^{(l)}$ in eqs. (\ref{e1})
and (\ref{e2}) are plagued by divergences due to the singular
behavior of the embodied 
perturbative QCD cross-sections $\hat{\sigma}^{(l)}(q_\perp^2)$ as the momentum 
transfer $q_\perp^2 \rightarrow 0$.

However, since the partons during the cascade evolution are generally
 off-shell, 
they are not massless but carry an invariant mass $|k^2| \ge max(\mu^2, m^2)$ 
where $\mu \approx 500$ MeV and $m$ is the flavor-dependent mass
  (equal to zero only for gluons).
This feature makes
it possible to employ
in VNI a {\it dynamical regularization} 
of the divergent pQCD cross-sections (rather than 
resorting to the common recipe of cutting out all parton interactions
involving momentum transfers $q_\perp^2 < p_0^2$ below some fixed $p_0$).
Combining the {\it uncertainty principle} 
$b_\perp q_\perp \, \lower3pt\hbox{$\buildrel <\over\sim$}\,1$
where $b_\perp$ and $q_\perp$ is the impact parameter
 and momentum transfer of two
colliding partons, respectively, with the {\it resolution requirement}
$q_\perp^2 \ge \max(k_1^2,k_2^2)$, where $k_{1,2}^2$ are the partons
invariant masses, one obtains
a distinct constraint $p_0 \rightarrow p_0(b_\perp,q_\perp) \ge p_0$
for each individual parton collision.
This procedure hence resolves the divergence problem,
except for the special case of gluons that are almost on-shell
and undergo a collision with very small $q_\perp$,
which are cut-off if  $q_\perp^2 < p_0^2$.
Specifically, we we use $p_0 = 1.1$ GeV for the present analysis,
as is the default value in VNI, 
which provides agreement with $pp$ data at $\sqrt{s}\approx 20$ GeV.
\footnote{
Notice that the applicability of pQCD down to $p_0=1.1$ GeV
may be justified by observing that  with $\alpha_s(p_0^2) \simeq 0.45$.}.

Although this dynamical regularization eliminates to large extent the
 arbitrariness
of a fixed cut-off $p_0=const$, it still ignores non-perturbative
contributions to parton-parton collisions. Neither
the detailed form, nor the significance of these latter
interactions, are known, notwithstanding their possible
impact on collective behavior under extreme high-temperature/density
conditions.
Therefore it may be worth while to attempt accounting for 
non-perturbative {\it soft} parton cascading ($p_\perp^2 < p_0^2$) in addition
to the truly perturbative {\it hard} parton evolution ($p_\perp^2 > p_0^2$),
and to study the consequences phenomenologically.
In the simulation program VNI, this option 
\footnote{
The default setting in VNI is {\it not} to include
soft parton collisions, but only the perturbatively calculable
hard interactions.}
is offered, and we have examined the non-perturbative effects by
comparing our results with and without inclusion of {\it soft} 
parton collisions.
The crucial observation was that
 switching off the soft-scatterings among the partons
resulted in no noticeable difference in the energy deposited in the system
or in any of the other observables examined in the present work.
This is easily understood, because the `soft' scatterings by
definition ($q_\perp^2 < p_0^2 = 1.1$ GeV) involve momentum transfers 
which are on the average a couple of hundred MeV only.
As a consequence, neither do these `soft' collisions generate
significant transverse energy, nor do they cause a notable
deflection of the involved partons - contrary to the `hard' collisions.

Thus, as the contribution of soft
scatterings is negligible at CERN SPS energy, the parton cascade development can be
described completely within pQCD without any further
model assumptions, i.e., by considering 
{\it hard} parton collisions only, and avoiding the phenomenological
inclusion of {\it soft} collisions.
\medskip

\subsection{Time-evolution of energy density and $E_\perp$-production}
\smallskip

The magnitude of transverse energy production and of the energy
density achieved in the central region of the nuclear collision
are  most crucial quantities for the detection of a QGP formation.
The transverse energy deposited in the collision can be used to estimate
the time-development of the  energy density in the
central collisions region, using the Bjorken relation~\cite{bj};
\begin{equation}
\epsilon(\tau) \;\approx\;  \frac{1}{\pi \overline{R}^2 \tau} 
\left(\frac{dE_T}{dy}\right)_{
{\scriptsize
\begin{array}{c}
|y-y_{cm}| < 0.5 \\
|z| < 0.5 \;fm\
\end{array}
}
}
\;,
\label{eps}
\end{equation}
where $\overline{R}<R_A \simeq 7$ $fm$ for $Pb$.
The simulation with VNI permits us to obtain the dynamic evolution of 
this density from the
first parton collision, through the cascading stage with re-interactions
and gluon emission, the parton recombination to pre-hadronic clusters,
and the emergence of final hadrons and resonances.
The top panel of Fig.~2 shows
the evolution of the total transverse energy  and also
the energy density resident in partons
and clusters and hadrons, with the proper time $\tau$ in the central
rapidity region.  We see a rapid build-up of the transverse energy carried
by the partons, which reaches a peak around 0.7 $fm$, and then starts
dropping as more and partons form clusters and hadrons. 
The bottom panel displays the evolution of the energy density (\ref{eps})
contributed by the partons and the clusters also provides a beautiful insight.
The energy density is rather large 
($\, \lower3pt\hbox{$\buildrel >\over\sim$}\,5$ GeV/fm$^3$) at very early 
times,  when the primary hard scatterings take place. After 
about 0.2 $fm$,  the increase
in the transverse energy deposited due to scatterings and radiations is
matched by the increase in the co-moving four-volume $\pi R^2 \tau \Delta y$,
where $R$ is the transverse size of the volume and the energy density will
not change. Once the partonic processes stop depositing transverse energy
in the system, the energy density carried by the partons will start decreasing,
due to the increase in volume. As more and more partons coalesce into
clusters or hadrons, the energy density of the partonic matter decreases
and that of the hadronic matter increases. Note the long period of time
over which the hadronic matter is created. Note also that the energy density
contained by the hadronic matter never exceeds 0.5 GeV/fm$^3$. This analysis
suggests the existence of partonic matter for a period of
almost 7 $fm$. Note that in order to get the energy density of the matter
(partonic or otherwise) at $\tau$ =1 $fm$ we have to weight the respective
densities with the volume fractions occupied by the partonic and the hadronic 
matter. Such details will be the subject of  a future publication.
\medskip

Can we can trust these results and conclusions? 
We believe that if the parton cascade model provides
a reasonable description to the data which have already started arriving
from the measurements of $Pb+Pb$ collisions at the CERN SPS, then
we can be very confident about the above discussion.
Indeed, as we shall address now, we find very decent agreement with data 
from the CERN SPS.
\medskip

\subsection{Comparison with CERN SPS data}
\smallskip

In Fig.~3 we have plotted our results for the transverse energy
distributions, averaged over the appropriate impact parameter range, which
corresponds to 2\% $\sigma_{\rm {min.~bias}}$.
The experimental data are taken from the measurements reported
by the NA49~\cite{na49_prl} and the WA98~\cite{wa98_qm96} collaborations.
A nice description is obtained without any adjustments of parameters.

The rapidity distribution of negatively charged hadrons are considered
to provide a sensitive test of the particle production mechanism in such
collisions, and we give our results  in Fig.~4. They are seen to
be in good agreement with the preliminary data. 

An enhanced production of strange particles has long been considered an
important manifestation of the quark-hadron phase transition. In the  PCM
strange quarks are produced from  flavor creation as well as flavor
excitation processes~\cite{klaus_str}. As the final-state interactions among the
produced hadrons 
is not yet implemented (see however Ref. \cite{ron}),
there is no additional production of strangeness, after the hadronization.
In view of this, the results shown in
Fig.~5 are most interesting, as we again find a quantitative agreement with
the data obtained on the rapidity distribution of strange mesons.

The transverse momentum distribution of hadrons are considered to provide
information about the temperature at the time of freeze-out and the flow
velocity, if any, of the matter produced in such collisions. In view of the
absence of final state interaction among the hadrons in the PCM used here,
we find reasonable description of the transverse mass distribution of
negative hadrons (Fig.~6) and of neutral pions (Fig.~7). The results
in both the figures are normalized at 0.9 GeV. In a recent paper
it has been shown that this agreement is improved considerably after a final
state hadron interaction is introduced~\cite{ron}. Thes results confirm that
the hadrons acquire a large part of their transverse momenta at the time of 
hadronization itself. The deviation of our results from the final data
(when available) will help to provide a quantitative estimate of the
importance of the final state interaction among hadrons.
\bigskip
\bigskip

\section{SUMMARY}
\bigskip

We have analysed $Pb+Pb$ collisions at the
CERN SPS with beam energy per nucleon 158 GeV ($\sqrt{s}/A = 17$ GeV)
by performing Monte Carlo simulations
within the framework of the  parton-cascade/cluster-hadronization
model PCM that involves
the dynamical interplay between parton production, evolution,
parton-cluster formation, and hadron production throug cluster decay.
We see this study as a step to 
describe high-energy nuclear collisions on
the microscopical level of the
space-time history of parton and hadron degrees of freedom
based on QCD and supplemented by phenomenology.
Our conclusions from this study are as follows:
\begin{description}
\item{(i)}
The simulations of $Pb+Pb$ collsions give
a very good overall description of the CERN SPS data on transverse energy distribution,
the multiplicity distribution, including those for $K$ mesons, and a good
description of the transverse momentum distribution of hadrons.
This success may be taken as a `post-hum' justification
of the applicability of a pQCD parton decription even at CERN SPS energies
- especially in view of the fact that neither 
additional model assumptions, nor tuning of parameters were involved.
\item{(ii)}
Contrary to wide-spread belief,
at CERN SPS energy
the `hard' (perturbative)  parton production and parton cascading 
is an important element for particle production
at central rapidities - at least in $Pb+Pb$ collisions. In effect 
it provides almost 50 $\%$ of final particles around mid-rapidity.
`Soft' (non-perturbative)  parton interactions
on the other hand are found to be insignificant in their
effect on final-state particle distibutions.
\item{(iii)}
The maximum achieved  energy density in the partonic
phase can reach $\epsilon\, \lower3pt\hbox{$\buildrel >\over\sim$}\,5$ GeV/fm$^3$
in the central region, thereby providing favorable conditions for a 
partonic plasma formation.
It drops below a `critical' density $\epsilon_c \approx 2$ GeV/fm$^3$ only after
about 2 $fm$.
On the other hand, the energy density of produced hadrons is always
substantially lower than 1 GeV/fm$^3$, making a hadronic plasma formation unlikely.
\end{description}
Especially the latter two points would have important consequencees:
If both these effects indeed turn out  so drastic at RHIC 
as compared to  CERN SPS, then this would
imply very   favorable conditions for the formation of a 
baryon-free quark-gluon plasma in the central rapidity region,
and possibly even a baryon-rich plasma in the nucleon-dense 
fragmentation regions.
\smallskip

We see these encouraging results as a motivation
to extend and deepen this study by looking
for instance at nuclear collisions with  other beams and different energy, or
at proton-nucleus collisions. We intend to pursue this project in the
near future.
In perspective of heavy-ion collisions at the BNL RHIC or  the CERN LHC,
the advocated PCM decription should provide a smooth extrapolation to
these future nuclear collisions beyond the CERN SPS,
since the relative importance of 
partonic and hadronic degrees are regulated by the 
multi-particle dynamics itself.
\bigskip
\bigskip
\bigskip
\bigskip

\section*{ACKNOWLEDGEMENTS}
We acknowledge useful comments by  Miklos Gyulassy, Ron Longrace and Bikash Sinha. 
This work was supported in part by the D.O.E under contract no.
DE-AC02-76H00016.

\newpage

\begin{figure}[t]
\centerline{ \psfig{figure=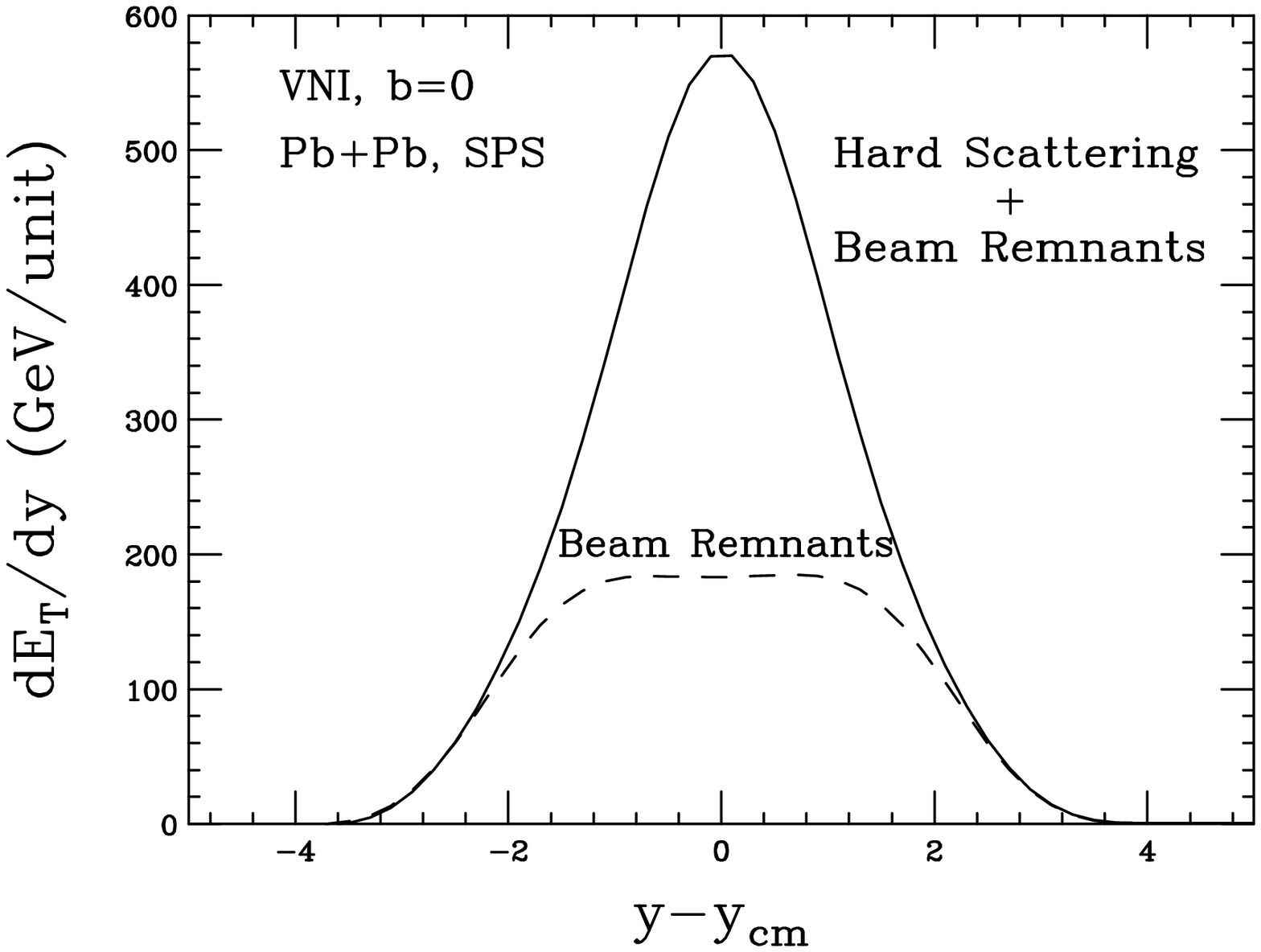,height=120mm}}
\caption{Transverse energy distribution in central collision of lead
nuclei at CERN SPS. The dashed curve gives the contribution of only the
remnant partons, which have not undergone scattering, while the
solid curve shows the sum of contributions from
parton cascade {\it and} the fragmentation of beam remnants. }
\end{figure}

\newpage

\begin{figure}
\epsfxsize=450pt
\centerline{ \epsfbox{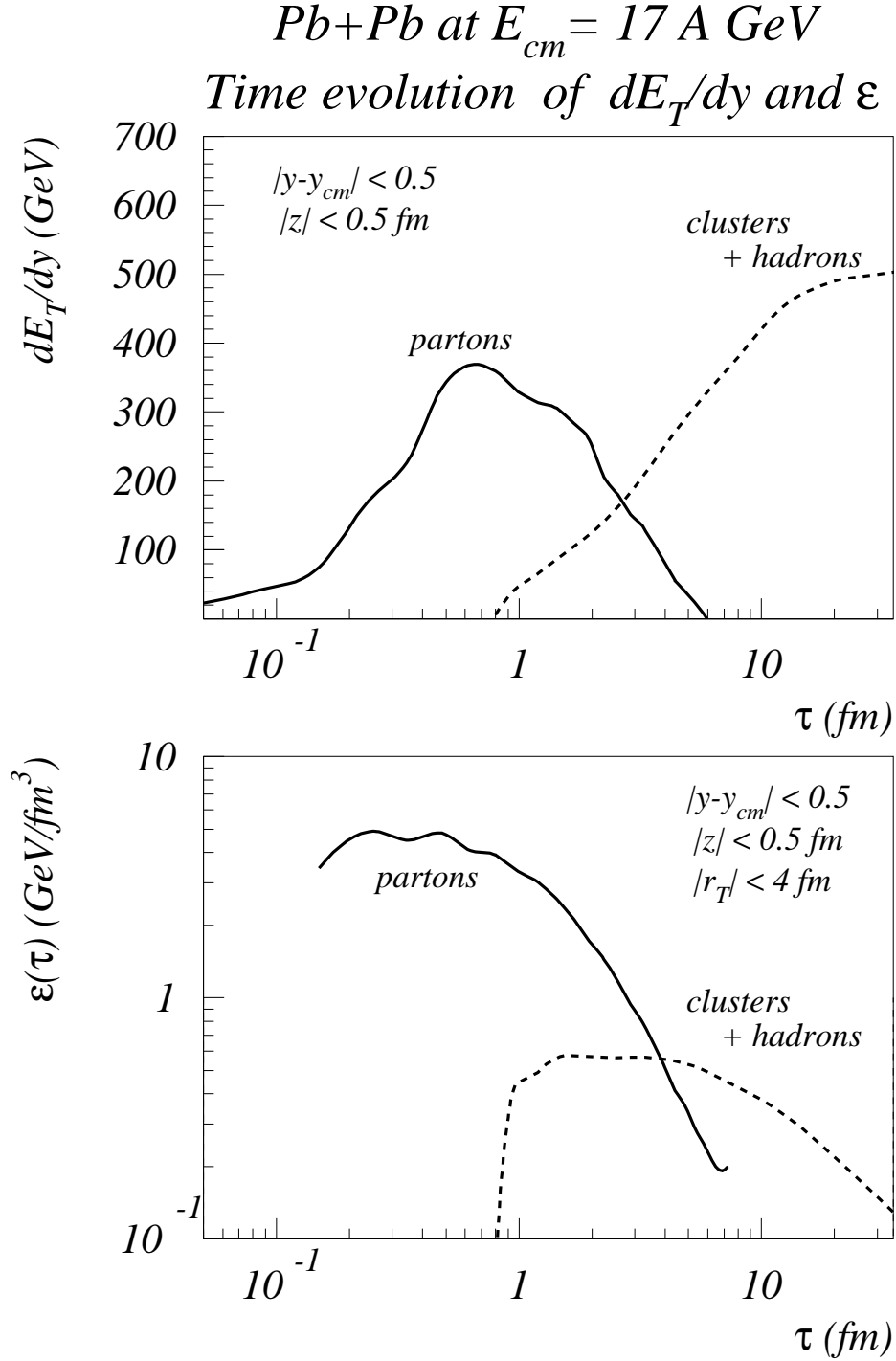} }
\vspace{-2.0cm}
\caption{ The evolution of the energy density of the partonic matter
in the central slice of the colliding lead nuclei at SPS energies.}
\end{figure}

\newpage

\begin{figure}[t]
\centerline{ \psfig{figure=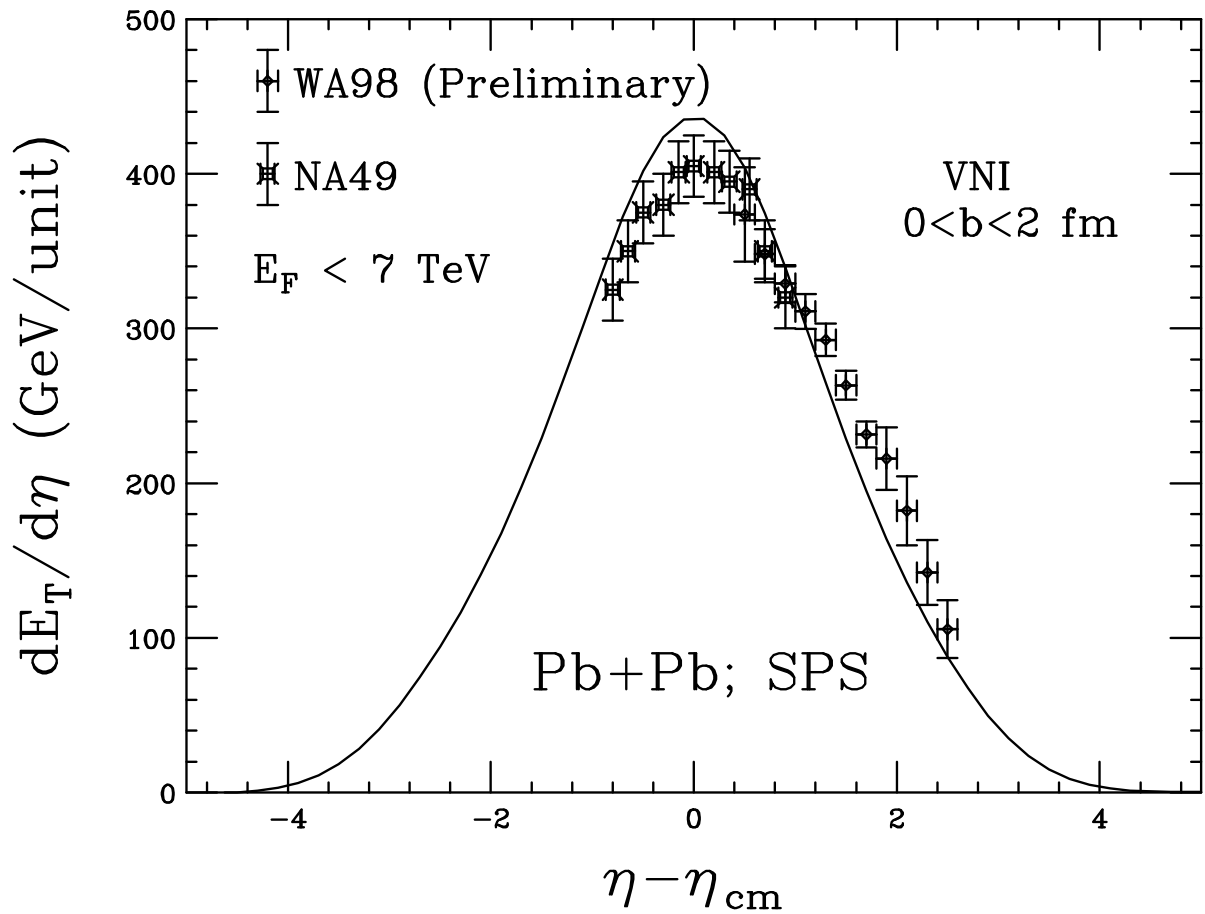,height=120mm} }
\caption{Transverse energy distribution in central collision of lead
nuclei at CERN SPS.}
\end{figure}

\newpage

\begin{figure}{t}
\centerline{ \psfig{figure=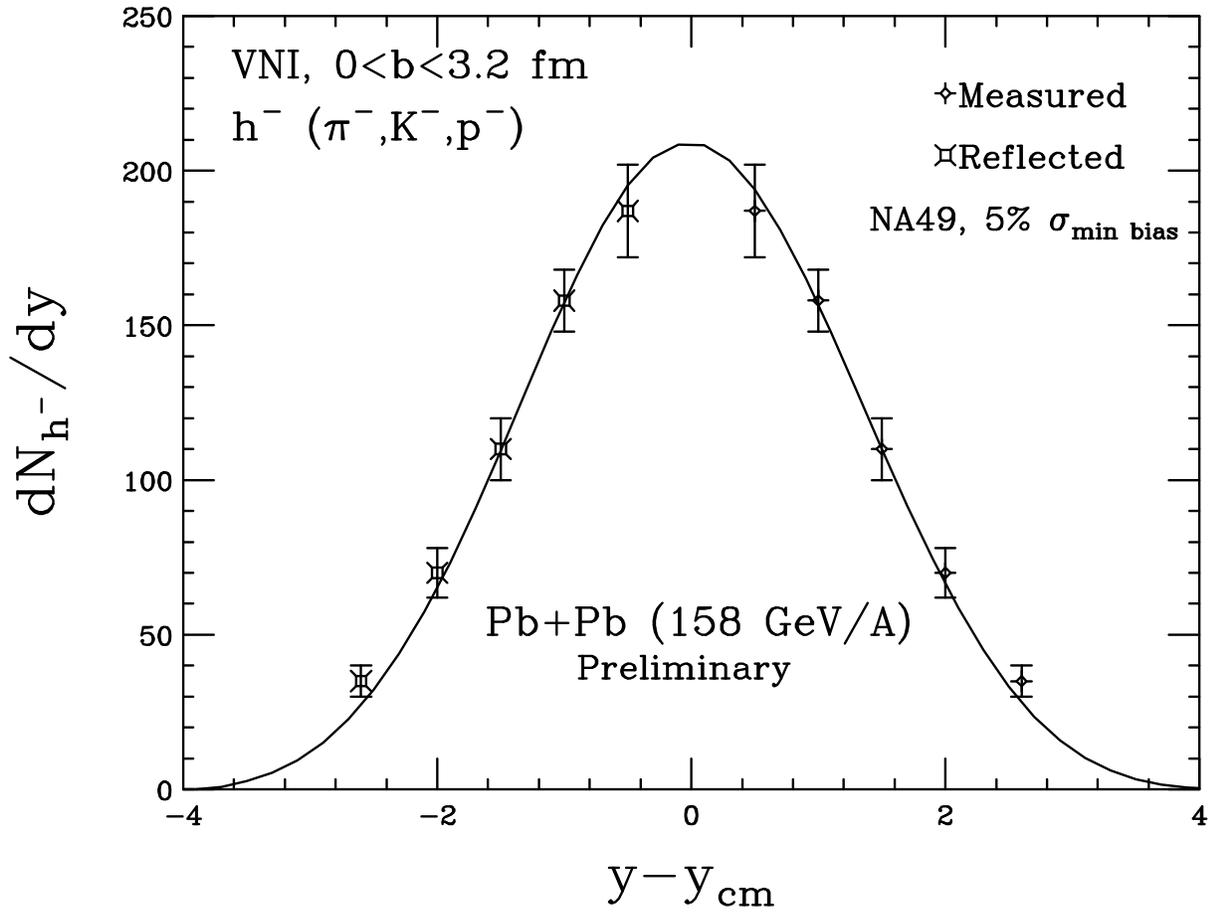,height=120mm} }
\caption{ The rapidity density distribution of negatively charged hadrons
($\pi^-$, $K^-$, and $\overline{p}$) in
 central collisions of lead nuclei at CERN SPS.}
\end{figure}
\newpage

\begin{figure}{t}
\centerline{ \psfig{figure=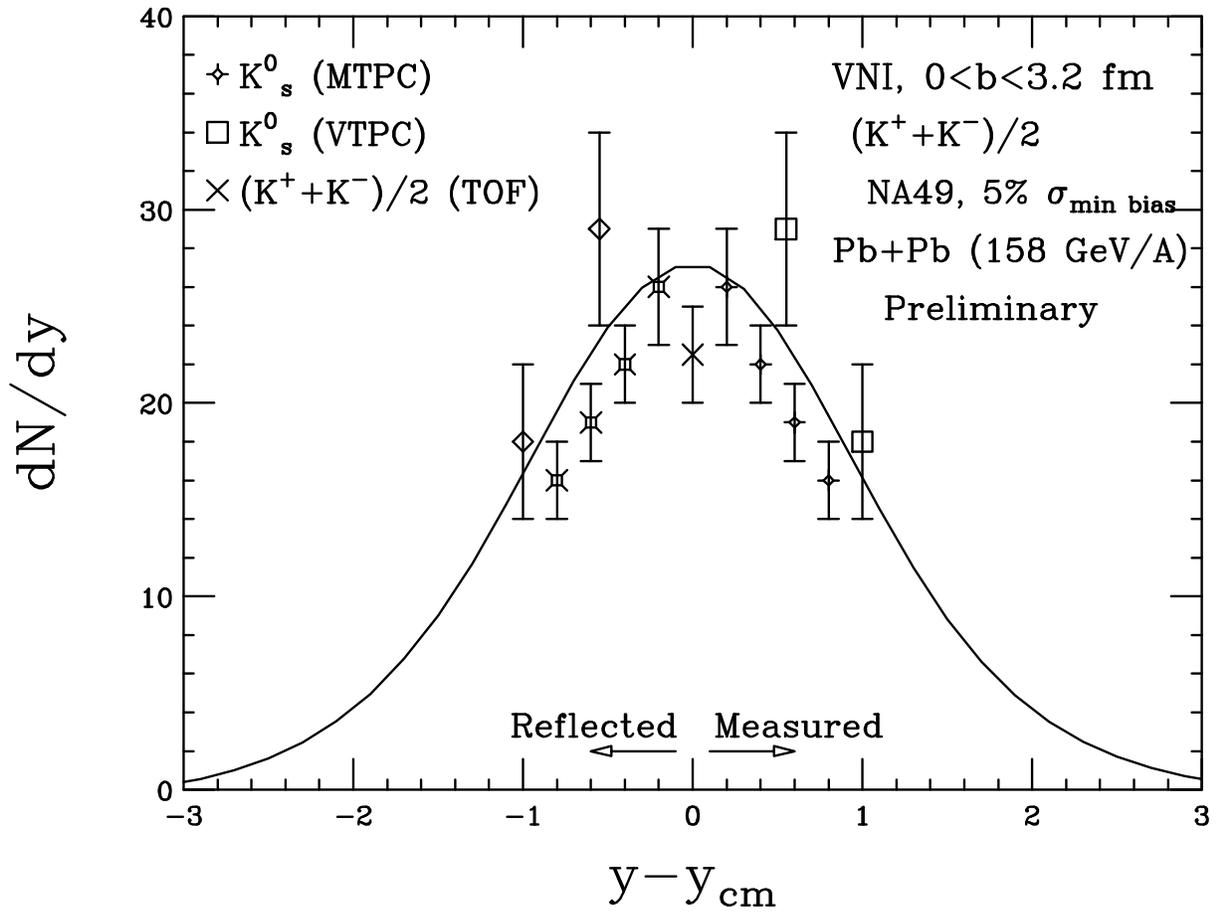,height=120mm} }
\caption{The rapidity density distribution of K mesons in central
collisions of lead nuclei at CERN SPS.}
\end{figure}
\newpage

\begin{figure}{t}
\centerline{ \psfig{figure=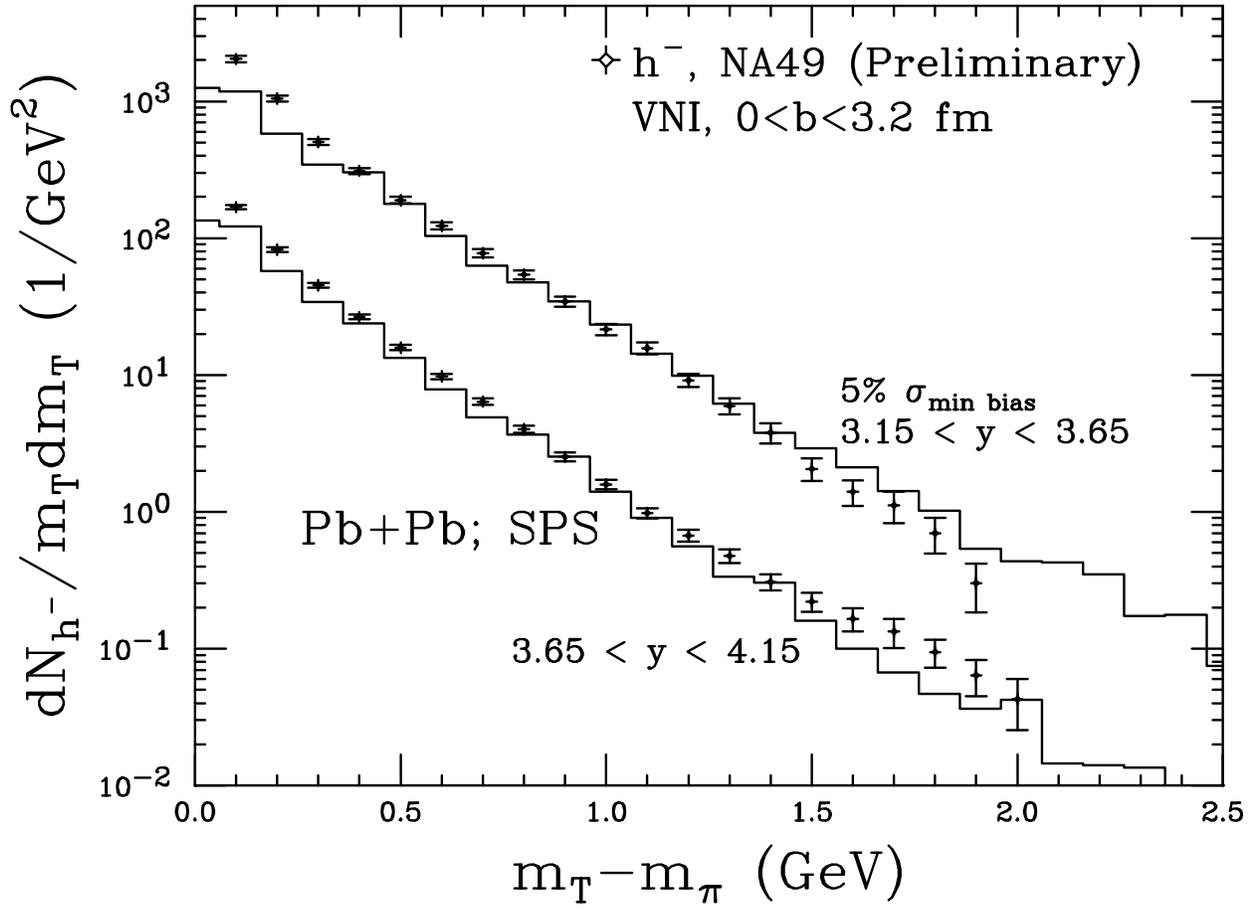,height=120mm} }
\caption{The transverse mass spectra of negative hadrons 
($\pi^-$, $K^-$, and $\overline{p}$)
successively scaled down by an order of magnitude. The predictions are 
normalized to the data ar $m_T-m_\pi$= 0.9 GeV}
\end{figure}
\newpage

\begin{figure}{t}
\centerline{ \psfig{figure=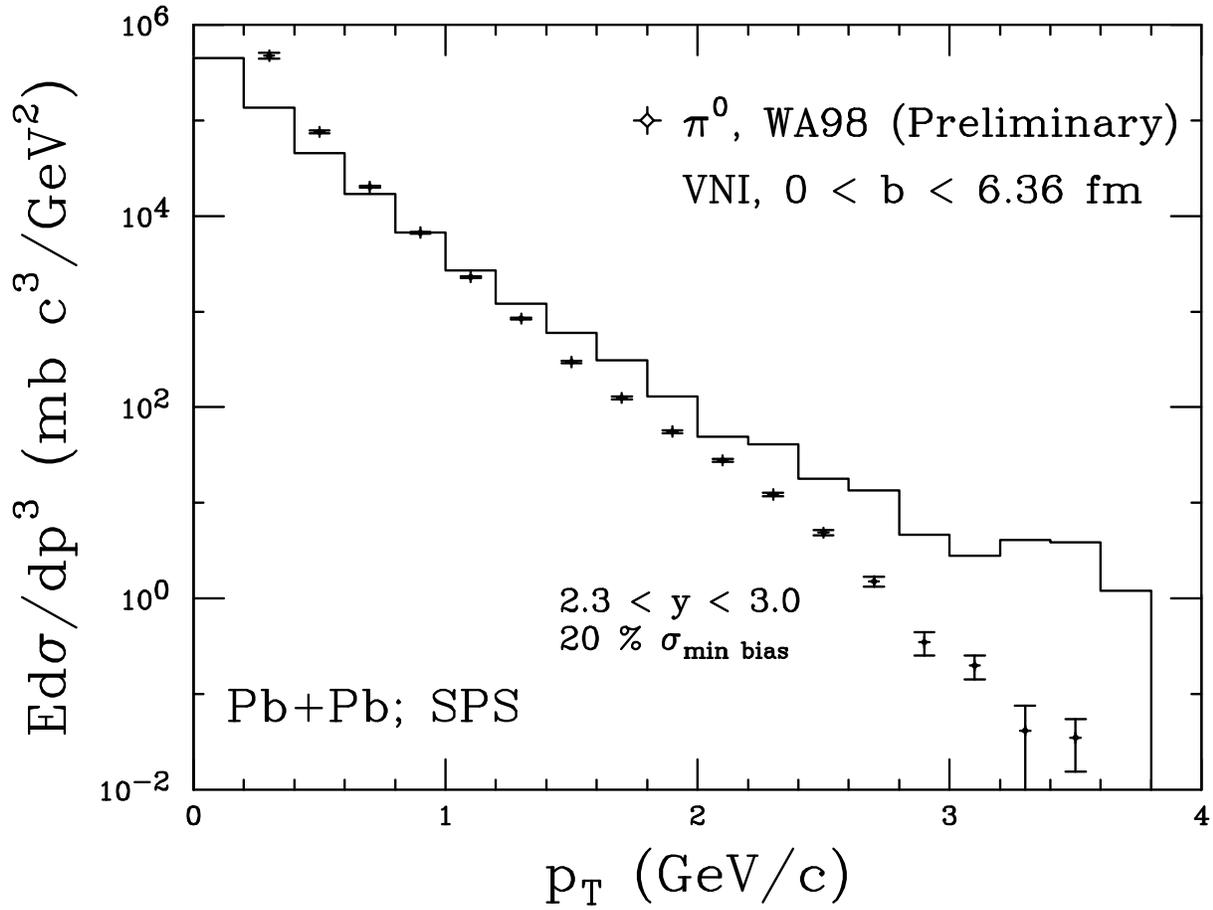,height=120mm} }
\caption{The transverse momentum distribution of neutral pions from central
collisions of lead nuclei at the CERN SPS. The predictions are normalized
to the data at 0.9 GeV}
\end{figure}

\end{document}